\documentclass[a4paper, amsfonts, amssymb, amsmath, reprint, showkeys, nofootinbib,superscriptaddress, twoside,aps,prd,longbibliography]{revtex4-1}
\pdfoutput=1
\usepackage[english]{babel}
\usepackage[utf8]{inputenc}
\usepackage[colorinlistoftodos, color=green!40, prependcaption]{todonotes}
\usepackage{amsthm}
\usepackage{mathtools}
\usepackage{physics}
\usepackage{xcolor}
\usepackage{graphicx}
\usepackage[left=23mm,right=13mm,top=35mm,columnsep=15pt]{geometry} 
\usepackage{adjustbox}
\usepackage{placeins}
\usepackage[T1]{fontenc}
\usepackage{lipsum}
\usepackage{csquotes}
\usepackage{hyperref}
\hypersetup{  
    colorlinks=true,                
    breaklinks=true,                
    urlcolor= black,                
    linkcolor= black,                
    bookmarksopen=false,
    filecolor=black,
    citecolor=black,
    linkbordercolor=black
}
\usepackage{units}
\usepackage{soul}
\usepackage{color}
\usepackage{changes}
\usepackage{subcaption}
\usepackage{diagbox}
\usepackage{stix}
\usepackage{ragged2e}
\usepackage{caption}
\hypersetup{pdfauthor={A. Jaroš, J. Toyfl, A. Pupić, B. Czasch, G. Boero, I.C. Bicket, P. Haslinger}, pdftitle={Electron Spin Resonance Spectroscopy in a Transmission Electron Microscope}}

\begin{document}
    \title{Electron Spin Resonance Spectroscopy in a Transmission Electron Microscope}

\author{Antonín Jaroš}
    \affiliation{Vienna Center for Quantum Science and Technology, Atominstitut, USTEM, Technische Universität Wien, Vienna, Austria}
\author{Johann Toyfl}
    \affiliation{Vienna Center for Quantum Science and Technology, Atominstitut, USTEM, Technische Universität Wien, Vienna, Austria}
\author{Andrea Pupić}
    \affiliation{Vienna Center for Quantum Science and Technology, Atominstitut, USTEM, Technische Universität Wien, Vienna, Austria}
\author{Benjamin Czasch}
    \affiliation{Vienna Center for Quantum Science and Technology, Atominstitut, USTEM, Technische Universität Wien, Vienna, Austria}
\author{Giovanni Boero}
    \affiliation{Institute of Electrical and Micro Engineering and Center for Quantum Science and Engineering École Polytechnique Fédérale de Lausanne, Lausanne, Switzerland}
\author{Isobel C. Bicket}
    \affiliation{Vienna Center for Quantum Science and Technology, Atominstitut, USTEM, Technische Universität Wien, Vienna, Austria}
\author{Philipp Haslinger}
    \email[]{philipp.haslinger@tuwien.ac.at}
    \affiliation{Vienna Center for Quantum Science and Technology, Atominstitut, USTEM, Technische Universität Wien, Vienna, Austria}

\date{\today} % Leave empty to omit a date

\begin{abstract}
Coherent spin resonance methods such as nuclear magnetic resonance (NMR) and electron spin resonance (ESR) spectroscopy have led to spectrally highly sensitive, non-invasive quantum imaging techniques with groundbreaking applications in fields such as medicine, biology, and physics. Meanwhile, transmission electron microscopy (TEM) offers detailed investigations with sub-atomic resolution, but often inflicts significant radiation damage. Here we exploit synergies and report on an integration of ESR spectroscopy in a TEM. Our miniaturized ESR setup on a standard TEM sample holder leverages the strong magnetic field of the TEM polepiece to align and energetically separate spin states. This integration will facilitate in-situ studies of spin systems and their dynamics, quantum materials, radicals, electrochemical reactions, and radiation damage — properties previously mainly invisible to electron microscopic tools. Moreover, this development marks a significant technological advancement toward microwave-controlled quantum spin studies with a highly controlled electron probe at the nanoscale.

%We report on an integration of ESR setup into TEM, for the first time ever bridging the gap between these resonant spectroscopies and electron microscopy. This innovation opens up new possibilities for GHz frequency sample excitation and signal detection within TEM, allowing in-situ investigations of the structure, dynamics, and distribution of paramagnetic species within the specimen. Spin state polarization is achieved with the magnetic field of the TEM polepiece, allowing for transition frequencies of up to 50 GHz. For driving the transitions, microresonator antenna is integrated into the TEM sample holder, together with a miniaturized field modulation coil for high spin sensitive measurements, resulting in sensitivity of 2.5 \textperiodcentered{} 10\textsuperscript{12 \textpm 1} spins/Hz\textsuperscript{1/2} similar to commercial ESR instruments. This represents an important technological advance towards microwave-driven spin studies with a highly controlled electron probe at the nanoscale, allowing for new in-situ studies of so far invisible properties to electron microscopic tools.

\end{abstract}

\keywords{Transmission Electron Microscope, ESR, NMR}

\maketitle

\newpage

%\section*{}

Nuclear magnetic resonance (NMR) and electron spin resonance (ESR) are non-invasive spectroscopic tools which rely on coherent microwave (MW) manipulation of spin states to determine the underlying chemical structure of substances. Together with electron microscopy (EM) \cite{knoll1932elektronenmikroskop}, which probes the specimen with highly energetic electrons to retrieve chemical and structural information at the atomic scale, these techniques are indispensable for studying entities at their most fundamental level. While ESR and NMR have achieved remarkable spatial resolution \cite{yon_solid-state_2017, lee_one_2001, geng_mapping_2021}, they still fall short by several orders of magnitude when compared to electron microscopic imaging techniques. Notable exceptions include specialized ESR setups, such as those using scanning tunneling microscopy or nitrogen-vacancy centers in diamond, which offer enhanced sensitivity but are limited to surface-level detection \cite{seifert_longitudinal_2020, baumann_electron_2015, simpson_ESRNV_2017}.

State of the art (scanning) transmission electron microscopy ((S)TEM), with the help of technological breakthroughs in aberration correction \cite{sawada2009stem47pm, Erni2009Atomic-resolution, hawkes2009AberrationHistory}, achieves imaging resolution of $<$45 pm using a complex system of electromagnetic lenses to shape the electron beam \cite{Ishikawa2023}. Despite recent advancements in cryo-electron microscopy, which have enabled high resolution imaging and tomography of biological specimens \cite{kuhlbrandt2014resolutionReview, yip2020ProteinTEM, Bai2015, nakane2020AtomicCryoEM}, radiation damage remains challenging to avoid, particularly in organic materials \cite{chen2020BeamSensitiveImaging, egerton2004RadiationDamage}.

Conversely, magnetic resonance spectroscopy/imaging \cite{Liang2000, Callaghan1993Principles} are widely applied spectroscopic tools based on coherent interaction with MWs for non-invasive investigation of nuclear and electron spin states with far-reaching applications in medicine\cite{Bullmore2012}, biology\cite{Borbat2001}, chemistry\cite{Weil1994, Wertz2012}, physics \cite{Ramsey1949}, and materials science \cite{Schweiger2001Principles}. ESR, for example, plays a pivotal role in investigating chemical reactions during electrolytic processes critical for battery research \cite{Nguyen2020, Zhao2021,Zhu2023}, polymer characterization \cite{uddin2020recent, xia2011epr} and dosimetry for the evaluation of radiation-caused damage\cite{regulla1982dosimetry}.

\begin{figure*}[t]
	\centering
	\includegraphics[]{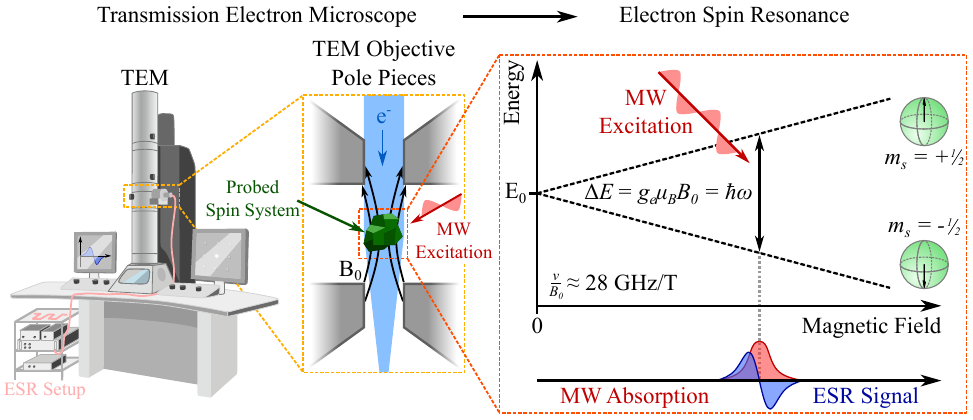}
	\caption{
 \justifying Overview of an ESR experiment in-situ in a TEM. A specimen containing addressable spin-1/2 particles is placed in the $B_0$ magnetic field generated by the objective lens polepieces of the TEM. The $B_0$ field induces Zeeman splitting in the energy levels of the spin system, as depicted on the right side. Transitions between the Zeeman levels are induced by an electromagnetic field of the appropriate frequency $\nicefrac{\nu}{B_0} \approx 28$ GHz/T, resulting in MW absorption and ESR signal.}	\label{fig:intro} 
\end{figure*} 

Combining ESR/NMR and TEM techniques and their complementary advantages will open up new avenues for research in these fields. Here we report on the first steps towards coherent spin manipulation within a TEM, bridging the fields of TEM and ESR/NMR \cite{Haslinger2024} and enabling correlative measurements between these two techniques - ESR-electron microscopy (ESR-EM), see schematics in Fig. \ref{fig:intro}. 

The probed spin system can be positioned on a custom built sample holder and placed inside a TEM. Within a magnetic field, spin-1/2 particles (such as electrons or protons) exhibit two distinct energy levels. For electrons, these energy levels are given by $E = m_s g_e \mu_B B_0$, where $m_s$ is the spin quantum number, $g_e$ is the electron g-factor, $\mu_B$ is the Bohr magneton, and $B_0$ is the magnetic field. The ground state ($m_s = -\nicefrac{1}{2}$) aligns anti-parallel with the field, while the excited state ($m_s = +\nicefrac{1}{2}$) aligns parallel, see Fig. \ref{fig:intro}. The energy separation $\Delta E=g_e \mu_B B_0$ of the spin states is described by the Zeeman effect. The strong and freely adjustable magnetic field within the TEM polepiece (0 to $\sim$1.8 T) at the specimen region provides the needed $B_0$ field for spin polarization. Transitions between the energy levels can be driven by MW fields at a resonance frequency of $\sim$28 GHz/T.

Nevertheless, integrating the required ESR setup into the limited space available in a TEM presents significant challenges. Between the upper and lower objective polepieces, there are only a few mm of available space for additional MW equipment. Furthermore, the samples and all MW components must be high vacuum-compatible.

\section*{Implementation}

\begin{figure*}[ht]
    \centering
    \includegraphics[]{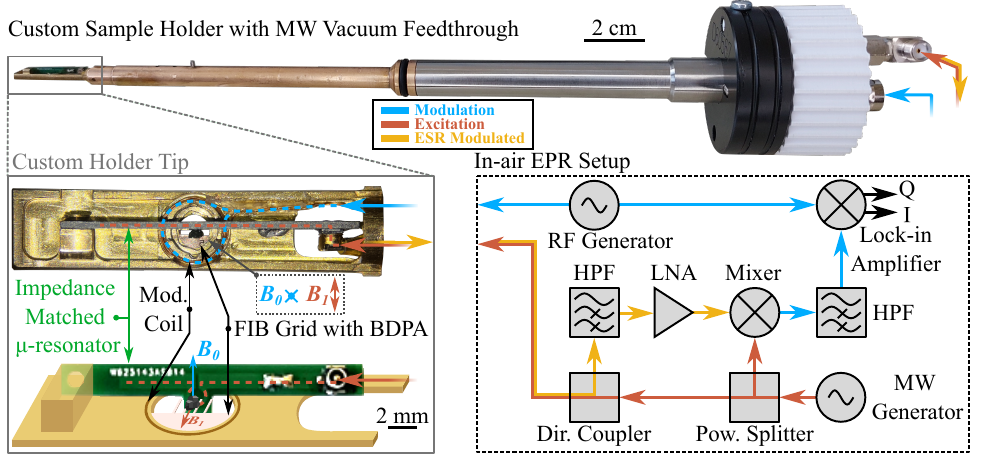}
    \caption{
    \justifying Custom TEM sample holder for ESR-EM, and experimental configuration. Microwaves ($\sim$4.5 GHz) and RF modulation signal (101 kHz) for the lock-in amplification scheme are fed into the modified standard TEM sample holder from an ESR setup outside the TEM. The impedance matched MW microresonator mounted on the sample holder tip generates an in-plane $B_1$ magnetic field, efficiently exciting a BDPA sample, which is positioned on a FIB lift-out grid in close proximity to the coil center. The modulation coil varies the out-of-plane $B_0$ magnetic bias field of the TEM objective lens. $B_1$ and $B_0$ are orthogonal to each other, see image insets. Back-reflected MWs ($\sim$4.5 GHz) together with the modulated ESR signal from the specimen ($\sim$4.5 GHz modulated with 101 kHz) are guided out of the TEM sample holder, decoupled from the main MW path, filtered with a high pass filter (HPF), amplified with low-noise amplifier (LNA), mixed down to the modulation frequency (101 kHz), and detected via a lock-in amplifier.}
    \label{fig:Setup}
\end{figure*}

In order to record ESR spectra in TEM, we implemented a miniaturized microwave circuit \cite{Boero2003, narkowicz2005planar, narkowicz2008scaling} on a customized TEM sample holder (designed for an FEI Tecnai F20 TEM), see Fig. \ref{fig:Setup}. This was done using a custom designed printed circuit board (PCB) with an $\Omega$-shaped microresonator, impedance matched to $\sim$4.5 GHz to enable effective excitation and readout of the spin states. This frequency range offers a wide variety of readily available high-level MW components. The sample is positioned on a FIB lift-out grid near the center of the microcoil to optimize coupling efficiency to the resonator while also allowing the necessary free space access for the electron beam during sample characterization.

While commercial ESR spectrometers with cavity design are mainly optimized for sample volumes of 1 mm$^3$ to 1 cm$^3$, with spin sensitivities of $10^9 - 10^{11}$ spins/Hz\textsuperscript{1/2}, such microcoil geometries have shown outstanding sensitivities down to $\approx 10^8$ spins/Hz\textsuperscript{1/2} at 300 K and 50 GHz for $<1$ mm$^3$ samples \cite{Boero2003, matheoud2017single}.

Following the derivations in the Methods section, we estimate the spin sensitivity of our microcoil, similarly to \cite{Boero2003}, by
\begin{equation}\label{eq: spin sens}
    N_{\text{min}} = \frac{NV_s}{\text{SNR} \sqrt{\Delta f}} = \frac{24 k_B^{3/2}}{ \gamma^3 \hbar^2} \frac{T^{3/2} \sqrt{R}}{B_u B_0^2} \ .
\end{equation}
Table \ref{tab:spin sens} presents our estimates of the spin sensitivity across a range of feasible parameters in the TEM.

\setlength\extrarowheight{3pt}
\begin{table}
    \centering
    \begin{tabular}{|c|c|c|c|}\hline
       \multicolumn{4}{|c|}{\textbf{Spin Sensitivity $N_{\text{min}}$} (spins/Hz\textsuperscript{1/2}) }\\
        \hline
        \backslashbox{\textbf{Bias Field $B_0$}}{\textbf{Temperature}} &  300 K &  77 K &  10 K  \\ 
        % 50 Ohm
        % \hline
        % 0.17 T $\hat{\approx}$ 4.8 GHz & $\sim 2\cdot 10^{10}$& $\sim 3\cdot 10^9$&$\sim 1\cdot 10^8$\\
        % \hline
        % 0.71 T $\hat{\approx}$ 19.9 GHz &$\sim 1\cdot 10^9$ & $\sim 2\cdot 10^8$&$\sim 7\cdot 10^6$\\
        % \hline
        % 1.8 T $\hat{\approx}$ 50.4 GHz & $\sim 2\cdot 10^8$& $  \sim 2\cdot 10^7$&$\sim 1\cdot 10^6$\\
        % \hline
        % 1 Ohm
        \hline
        0.17 T $\hat{\approx}$ 4.8 GHz & $\sim 2.9\cdot 10^{9}$& $\sim 3.8\cdot 10^8$&$\sim 1.8\cdot 10^7$\\
        \hline
        0.71 T $\hat{\approx}$ 19.9 GHz &$\sim 1.7\cdot 10^8$ & $\sim 2.2\cdot 10^7$&$\sim 1.0\cdot 10^6$\\
        \hline
        1.8 T $\hat{\approx}$ 50.4 GHz & $\sim 2.6\cdot 10^7$& $  \sim 3.4\cdot 10^6$&$\sim 1.6\cdot 10^5$\\
        \hline
    \end{tabular} 
    \caption{ \justifying Estimated theoretical spin sensitivity for the ESR-TEM sample holder using an $\Omega$-shaped microresonator, projected to various combinations of bias magnetic field $B_0$ and sample temperatures. We estimate the spin sensitivity using Eq.~\ref{eq: spin sens}, assuming a resistance of $R = 1 \ \Omega$ due to the ohmic resistance of the coil and a coil diameter of $d = 1 $ mm.}
    \label{tab:spin sens}
\end{table}

To target the operating frequency $\nu = 4.5$ GHz, we operate the TEM in low magnification mode, which allows us to produce a $B_0$ field between approximately 0 to 0.8 T at the specimen. We tune the objective lens strength to achieve a field of $B_0 = 160$ mT, matching the transition frequency $\nu = 4.5$ GHz. Low magnification mode serves a dual purpose: facilitating spin state polarization for ESR measurement and enabling extended camera lengths in diffraction mode for the observation of small electron beam deflections, paving the way for future investigations with a highly controlled electron probe at the nanoscale \cite{Haslinger2024}.

\begin{figure*}[ht]
    \centering
    \includegraphics[width=1\textwidth]{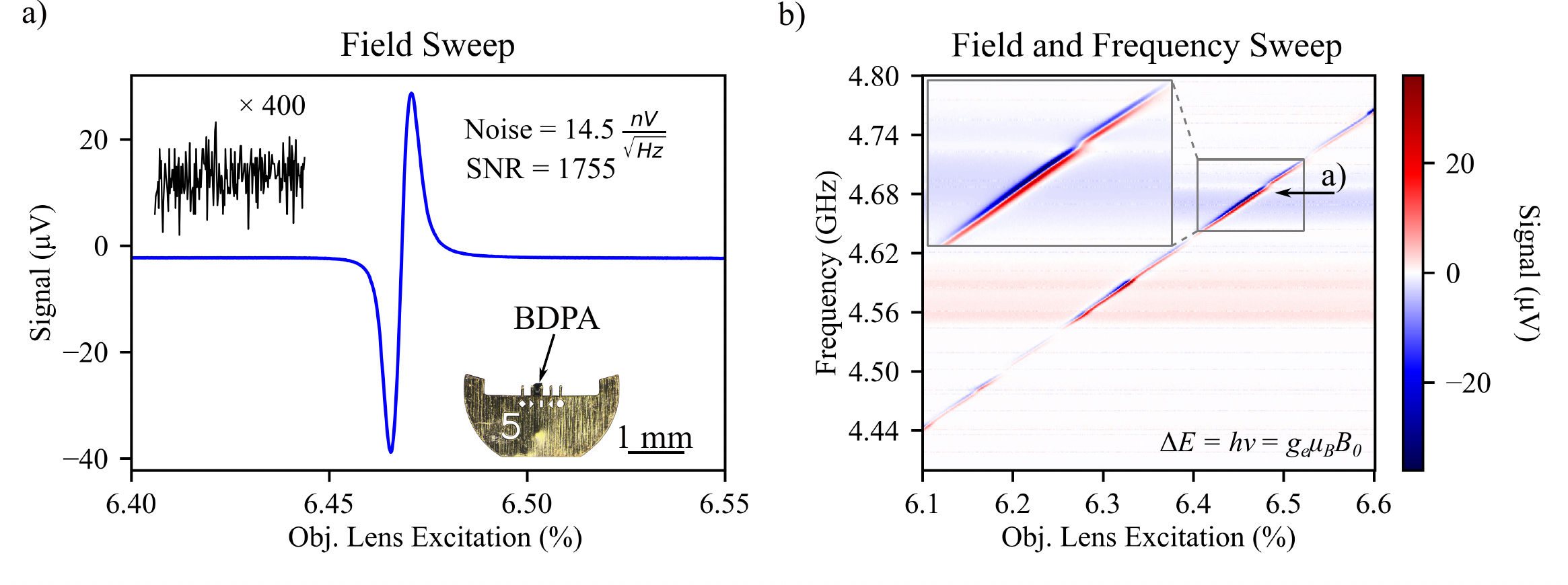}
    \caption{
    \justifying ESR spectra of a BDPA specimen measured in-situ in the TEM on the custom-built sample holder. In (a), a field sweep of the TEM objective lens is presented. The insets of the figure show the noise floor of our measurement and the \textmu m-sized BDPA specimen positioned on a FIB lift-out grid. Comparing the rms value of the noise and peak-to-peak signal amplitude leads to an SNR $ \approx 1700$ and a spin sensitivity $N_{min} \approx 3 \cdot 10^{12}$ spins/Hz\textsuperscript{1/2}. In (b), a combined frequency and objective lens excitation sweep is presented, demonstrating broadband frequency ESR detection capabilities. The ESR signal follows a straight line, consistent with the expected Zeeman energy-level splitting. The intensity variations of the ESR signal at different excitations come from the microresonator impedance match. Additional shifts may also occur due to the effects of phase changes across the impedance match. The DC offsets can be caused by mechanical vibrations of the holder induced by the modulation coil, leading to effective changes in the reflected signal. Both (a) and (b) were measured in an FEI Tecnai F20 TEM. Experimental settings are as follows: MW generator output power $P_{g} = 20$ dBm; lock-in time constant $\tau = 100$ ms (a) and $\tau = 30$ ms (b).}
    \label{fig:ESR-comp}
\end{figure*}

The ESR setup with lock-in amplifier readout is similar to the one described in \cite{Boero2003}. It employs a Rohde\&Schwarz SMB100B MW generator to drive transitions between the specimen's spin states. The MWs ($\sim$4.5 GHz) are guided from the in-air MW equipment to the tip of the TEM sample holder in vacuum via a custom vacuum feedthrough. For low-noise ESR investigations, a miniaturized modulation coil on the sample holder modulates the $B_0$ field. In turn, the ESR signal originating from the sample is modulated as well. The modulated ESR signal is then received by the MW microresonator and routed out of the TEM, decoupled from the mainline via a directional coupler, amplified, filtered, and down-mixed with the reference signal to the modulation frequency (101 kHz). Final data acquisition is performed on a lock-in amplifier (Stanford Resarch SR810), which generates in-phase (denoted as I) and quadrature signals (Q). When a lock-in detection scheme is used, only signals at the chosen modulation frequency can be detected, resulting in an increased signal to noise ratio. By sweeping the MW frequency or $B_0$ field across the resonance and simultaneously recording the I and Q channels, a continuous wave (CW) ESR spectrum is obtained, representing the first derivative of the ESR absorption signal. Moreover, by adjusting the MW phase difference on the mixer, it is also possible to record an ESR dispersion signal \cite{talpe1966dispersion}. Figure \ref{fig:Setup} shows a schematic of the ESR microwave setup and the TEM sample holder. 

\section*{In-situ ESR-EM}

To characterize our setup and its spin sensitivity $N_{min}$, we choose as our specimen the electron spin-active radical, $\mathrm{\alpha,\gamma}$-bisdiphenylene-$\mathrm{\beta}$-phenylallyl (BDPA) \cite{BOERO_BDPA}, which is widely used in ESR as a benchmark sample due to its stability at room temperature and high spin density, $\rho \approx 1.5 \cdot 10^{27}$ spins/m$^3$.

Fig. \ref{fig:ESR-comp}a shows a CW ESR spectrum of a $\sim$150$\times$150$\times$150 \textmu m$^3$-sized BDPA sample, positioned on a FIB lift-out grid. The blue curve represents the in-phase (I) signal. The quadrature (Q) signal is not depicted, as the reference modulation phase was tuned to merge all signal into the I channel. The MW phase difference at the first mixer was tuned to observe the absorptive ESR spectrum. The MW frequency was set to $\nu = 4.695$ GHz, while the objective lens excitation was swept from 6.4000\% to 6.5500\% (with a step size of 0.0002\%). The ESR resonance frequency follows $\nu = 0.64 E_{obj} + 0.536$, where $\nu$ is a frequency in GHz and $E_{obj}$ is the objective lens excitation in percentage. 

Additionally, as BDPA is a well-established calibration sample, we use it as a reference to calculate the corresponding $B_0$ field at the specimen. The relationship is given by $B_0 = 22.86 E_{obj} + 19.14$, where $B_0$ is the magnetic field in mT. An offset of approximately 19 mT is observed, attributed to field leakage from the minicondenser lens and remanent fields of the objective polepiece \cite{f20interfacemanual, f20modesmanual}. Based on this calibration, the 6.4000\% to 6.5500\% sweep corresponds to a $B_0$ field sweep from 165.43 mT to 168.86 mT. Each data point was measured with a time constant $\tau = 100$ ms. The output power of the MW generator was optimized for the highest signal-to-noise ratio (SNR) and set to $P_{g} = 20$ dBm. 

Given that the magnetic field of the polepieces exhibits gradients to tightly focus the electron beam onto the sample \cite{Reimer2008Transmission}, we anticipated a broadening of the ESR line. However, no significant broadening on the length scale of our microscale specimen was found for our measurements (peak-to-peak $\Delta \nu_{pp} \approx 3.2$ MHz, compared to the expected $\Delta \nu$ range of 1.5 to 2.9 MHz \cite{mitchell2011electron}). If sufficiently small samples are selected, such as those used in electron microscopy, the magnetic field changes across the sample are negligible, and do not lead to a broadening of the line width.

Furthermore, by precisely adjusting the excitation of the objective lens polepiece with a smallest possible step size of 0.0001\% (corresponding to $\Delta B_0 \approx 2$  \text{$\mu$}T and $\Delta \nu \approx 50$ kHz), the magnetic field $B_0$ can be finely tuned and used for precise ESR field sweeps, as $B_0$ is a linear function of the objective lens excitation. In general, magnetic field sweeps are commonly used by commercial ESR spectrometers instead of frequency sweeps due to constraints imposed by the resonator design. The MW resonator, due to its high Q-factor, is typically locked to a certain frequency. In contrast, our setup enables concurrent frequency and field sweeps, as represented in Fig. \ref{fig:ESR-comp}b, without significant changes in signal amplitude, due to a lower resonator Q-factor. Here, the MW frequency was swept from  $\nu = 4.4$ to 4.8 GHz in steps of  $\Delta \nu = 0.5$ MHz. The objective lens excitation was swept from 6.100\% ($\sim$158.6 mT) to 6.600\% ($\sim$170.0 mT), with a step size of 0.001\%. A linear shift of the ESR signal with the objective lens excitation is evident and consistent with the expected Zeeman energy-level splitting. The signal strength variations may originate from the microresonator impedance match, additional phase changes across the impedance match, and vibrations induced by the modulation coil. Each data point was measured with a time constant $\tau = 30$ ms.

We have reached an $\text{SNR} \approx 1700$ and a spin sensitivity $N_{min} \approx 3 \cdot 10^{12}$ spins/Hz\textsuperscript{1/2}. The background noise level of our measurement is shown in the inset of Fig. \ref{fig:ESR-comp}a. 
%For reference, we measured the sample using a commercial ESR spectrometer setup (Bruker EMS Spectrometer) at 9.5 GHz (data not shown). Resulting spin sensitivity was $N_{min} = 1 \cdot 10^{10}$ spins/Hz\textsuperscript{1/2}. 
The theoretical spin sensitivity evaluated by eq. \ref{eq: spin sens} is $N_{min} \approx 3 \cdot 10^{9}$ spins/Hz\textsuperscript{1/2}. The discrepancy between the theoretical and experimental value is mainly attributed to the use of non-optimized MW electronics and is discussed in the Methods. Optimizations of our electronic setup should increase the spin sensitivity towards achieving the theoretical maximum, as given in Table~\ref{tab:spin sens}. Overall, our setup allows effective ESR spectra measurements in-situ in the TEM. The same setup can be also used for ferromagnetic resonance experiments, which exhibit greater spin polarization and would yield a stronger signal.

\begin{figure}[t]
    \centering
    \includegraphics[width=0.48\textwidth]{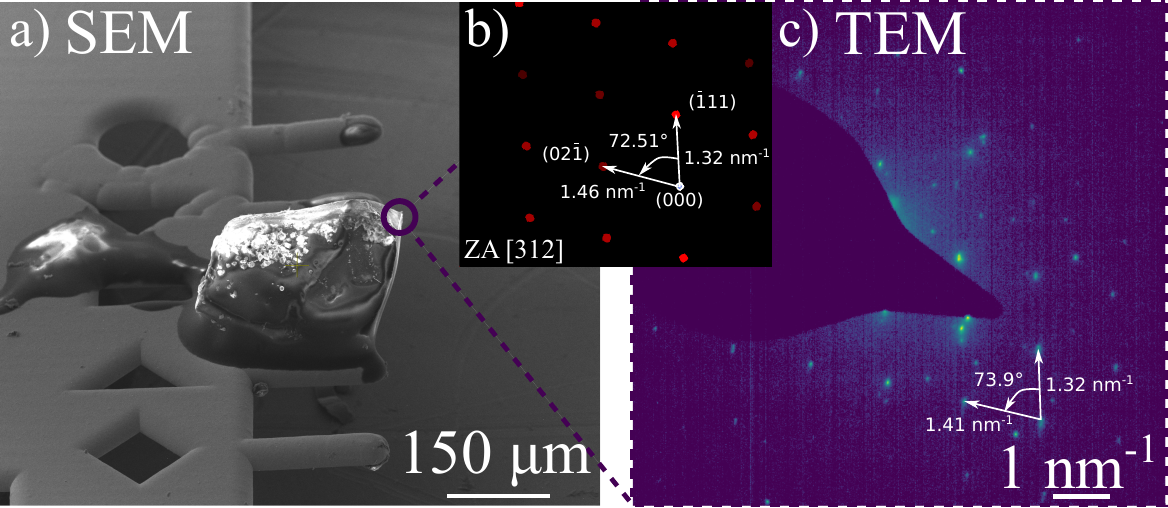}
    \caption{
    \justifying
    Secondary electron image of the BDPA crystal used for ESR measurements, glued on a FIB lift-out grid (a). Calculated electron diffraction pattern from the [312] BPDA zone axis (b) and experimental (c) TEM diffraction pattern from a location at the crystal's edge.}
    \label{fig:TEM_measurement}
\end{figure}

Furthermore, the sensitivity of our setup can be improved by increasing the bias magnetic field $B_0$ to $\sim$1.8 T, which is for standard usage the upper threshold for the FEI Tecnai F20 TEM, resulting in a transition frequency of $\nu_0=$ 50 GHz and a corresponding improvement, $\propto \nu_0^2$, of the SNR by a factor of $100\times$. Enabling operating conditions at cryogenic temperatures would also be beneficial, as are often used in cryo-TEM \cite{Zhu2021,Rennich2024}. A temperature reduction from room temperature to 77 K (liquid nitrogen) or 4 K (liquid helium) leads to a spin polarisation gain, $\propto T$, of $4\times$ and $75\times$, respectively, as well as significant improvement, $\propto T$\textsuperscript{1/2}, of the thermal readout noise (Johnson-Nyquist noise). Maximizing both values leads to a theoretical spin polarisation of $\sim35\%$ and a spin sensitivity of $1.8 \cdot 10^{5}$ spins/Hz\textsuperscript{1/2}, sufficient to investigate even nanometer-scale ESR samples.

Fig. \ref{fig:TEM_measurement}a shows a secondary electron microscope (SEM) image of our BPDA sample, providing morphological information elusive in ESR. We used the SEM image to estimate the sample volume for determining spin sensitivity. Although this secondary electron image was taken in a different instrument, several commercially available TEMs are also equipped with the necessary electron detectors \cite{zhu2009SE_TEM, mitchell_secondary_2016}. 

To demonstrate the possibilities for correlative ESR and TEM characterization of the same specimen under the same environmental conditions, we show in Fig. \ref{fig:TEM_measurement}c a diffraction pattern of the same BDPA crystal as in Fig. \ref{fig:ESR-comp}. Positioning a parallel 200 keV electron beam near the edge of the specimen, where the sample is thin enough to observe transmitted electrons, we use a Gatan Orius bright field camera to capture the diffraction pattern, with an acquisition time of 1 s. The image was filtered using a median filter to remove camera speckle. The pattern is dominated by reflections from a single zone axis, but the presence of other diffraction spots indicates the presence of multiple crystallites within the area illuminated by the electron beam. The lattice spacings and angles of the dominant zone axis observed are consistent with the diffraction pattern calculated along the $[312]$ zone axis (Fig.~\ref{fig:TEM_measurement}b). For the calculation, we construct a model of the crystal from the atomic coordinates given by Azuma \textit{et al.} \cite{azuma_molecular_1994} for BDPA-benzene and calculate the electron spot pattern using the electron microscopy simulation software JEMS (Electron Microscopy Software Java Version) v5.

From this correlative measurement, we can derive crystallographic information, such as the lattice spacing, of localized areas of the specimen. Both measurements in Fig. \ref{fig:TEM_measurement} are recorded after ESR spectroscopy by means of standard EM techniques. These measurements highlight the versatility of the custom-built holder and demonstrate the capability for simultaneous or sequential measurements with both the electron beam and our ESR-EM setup.

\section*{Conclusions}

In this study, we have successfully developed and implemented an electron spin resonance setup, tailored to microscopic sample sizes, in a transmission electron microscope. The ability to dynamically excite spin systems directly in the TEM ultimately leads to novel sensing capabilities of detecting localized spins using a highly controlled electron probe at the nanoscale \cite{Haslinger2024}. This extends the accessible spectral range and enables in-situ investigations of properties previously invisible to electron microscopy. 

The possibilities of ESR-EM not only open new horizons in various fields, such as in the study and fabrication of nitrogen-vacancy colour centres \cite{Mindarava2020,Shames2017,Dantelle2010}, archaeometry \cite{Rink1997}, battery research \cite{Nguyen2020, Zhao2021,Zhu2023}, and nanomagnetism and spintronics \cite{misra2024EPRSpintronics, slota2020MolecularSpintronics, harvey_ultrafast_2021}, but also simplify and improve workflows in disciplines that utilise both EM and ESR for analytical insights.

Further, this innovation enables precise investigations into the formation of radicals caused by electron-induced radiation damage, a fundamental challenge limiting the resolution and quality of electron microscopy analysis \cite{chen2020BeamSensitiveImaging, Flannigan2023, Ilett2020} of biological samples and in materials science \cite{ Yao2023, Watanabe2020}. Our setup could be used for assessing electron beam radiation damage \cite{regulla1982dosimetry}, by providing correlative ESR and TEM measurements inside the TEM.
%, paving the way for unprecedented in-situ studies of spin dynamics combined with the vast nanoscale characterization functionalities of TEM.

%This innovative integration not only opens up new possibilities for various fields, but also simplifies and improves workflows in disciplines that utilize both electron microscopy and ESR for analytical insights.

%$In addition, 

Spin resonance studies are at the forefront of non-invasive, coherent spectroscopic methods and offer a complementary perspective to traditional electron microscopy. This synergy is not only additive but transformative and promises to usher in a new era of coherent investigation of quantum systems with electrons at the nanoscale \cite{Haslinger2024, Ratzel2021, Gover2020free,ruimy2021toward,zhang2022quantum,kfir2021optical, Echternkamp2016Ramsey-type, morimoto2018diffraction, baum2013physics}.

\section*{Acknowledgments}
The authors thank Thomas Schachinger, Stefan Löffler, Michael Stöger-Pollach, Michael Eisterer and the ATI electronic and mechanical workshop for assistance in early stages of the experiment.
PH thanks the Austrian Science Fund (FWF): Y1121, P36041, P35953. This project was supported by the ESQ-Discovery Program 2019 "Quantum Klystron" hosted by the Austrian Academy of Sciences (ÖAW) and the FFG-project AQUTEM.

\bibliographystyle{naturenano}
\bibliography{main}

\section*{Methods: Spin sensitivity estimates}
To estimate the spin sensitivity of our MW setup, we follow \cite{Boero2003}. 
According to the principle of reciprocity \cite{hoult1976}, the voltage $\xi(t)$ induced into the microcoil by the sample can be expressed as
\begin{equation}
    \xi(t) = - \int_{V_s} \frac{d}{dt} \left( \boldsymbol{B}_{u}(\boldsymbol{r}) \cdot \boldsymbol{M}(t, \boldsymbol{r}) \right) \dd V \ ,
\end{equation}
where $\boldsymbol{B}_{u}(\boldsymbol{r})$ is the magnetic field generated by a unitary current (1 A) in the microcoil, $\boldsymbol{M}(t, \boldsymbol{r})$ is the varying magnetization of the sample and $V_s$ is the sample volume. Under quasi-steady state \cite{Weil1994} conditions, far from saturation and assuming that the unitary magnetic field $\boldsymbol{B}_{u}(\boldsymbol{r})$ and the sample magnetization $\boldsymbol{M}(t, \boldsymbol{r})$ is homogeneous in the sample volume and negligible elsewhere, the maximum voltage $\xi_{\text{max}}$ induced can be approximated by
\begin{equation}
    \xi_{\text{max}} \approx \gamma B_{1}M_{0} \left( \frac{B_{0}}{\Delta B} \right) B_{u}V_s \ ,
\end{equation}
where $\gamma$ is the gyromagnetic ratio of the electron, $B_1$ is the MW magnetic field, $M_0$ is the static magnetization, $B_0$ is the static polarizing magnetic field and $\Delta B$ is the half width at half maximum of the absorption line.

The static magnetization can be calculated by applying Curie's law \cite{kardar_statistical_2007} for a spin-1/2 system with Land\'{e} \textit{g}-factor, $g_e=2$
\begin{equation}
M_0 = \frac{N \gamma^2 \hbar^2 }{4 k_B T}B_0 \ ,
\end{equation}
where $N$ is the spin density of the sample (in spins/m$^3$), $k_B$ is the Boltzmann constant and $T$ is the sample temperature. Under ideal conditions, the maximum signal-to-noise ratio (SNR) is the ratio of the maximum signal to the thermal noise of the microcoil
\begin{equation}\label{eq: snr}
    \text{SNR} = \frac{\xi_{\text{max}}}{3 N_{\text{RMS}}} = \frac{\gamma B_{1}M_{0} \left( \frac{B_{0}}{\Delta B} \right) B_{u}V_s}{3 \sqrt{4 k_B T R \Delta f}} \ ,
\end{equation}
where $R$ is the noise equivalent resistance of the microcoil generating the thermal noise, $\Delta f$ is the equivalent noise bandwidth of the measurement electronics and the factor of 3 stems from the fact that one should be able to distinguish the signal from the noise at an SNR of 1.

The spin sensitivity in spins/Hz\textsuperscript{1/2} represents the minimum number of spins that are needed for detection in a sample. Assuming that one can increase $B_1$ up to $\Delta B$ without significantly broadening the linewidth due to saturation, the spin sensitivity can be calculated by
\begin{equation}\label{eq: spin sens methods}
    N_{\text{min}} = \frac{NV_s}{\text{SNR} \sqrt{\Delta f}} = \frac{24 k_B^{3/2}}{ \gamma^3 \hbar^2} \frac{T^{3/2} \sqrt{R}}{B_u B_0^2} \ .
\end{equation}
The unitary field of a one-turn microcoil is approximated by $B_u \approx {\mu_0}/{d}$, where $\mu_0$ is the permeability of the vacuum and $d$ is the coil diameter. The noise equivalent resistance of the PCB is estimated to be $1 \ \Omega$. Estimates of the spin sensitivity for various attainable parameters in the TEM, using eq. \ref{eq: spin sens methods}, are displayed in table \ref{tab:spin sens}.

The discrepancy between the estimated spin sensitivity from eq. \ref{eq: spin sens}, $N_{min} \approx 3 \cdot 10^{9}$ spins/Hz\textsuperscript{1/2}, and the measured value, $N_{min} \approx 3 \cdot 10^{12}$ spins/Hz\textsuperscript{1/2}, can be mostly explained by taking into account additional factors. The sample position was initially chosen to provide space between the FIB lift-out-grid and PCB for subsequent electron beam measurements (see Fig. \ref{fig:TEM_measurement}), which significantly decreased the sample-microcoil coupling. In our additional measurements, we positioned the BPDA sample directly in the center of the microcoil, improving the sensitivity to $N_{min} \approx 2 \cdot 10^{11}$ spins/Hz\textsuperscript{1/2}, but making the specimen less accessible for electron beam characterisation studies. Additionally, the ESR signal is reduced by the total noise factor of our MW setup (directional coupler: 10 dB, high pass filter:  0.5 dB, LNA: 1.2 dB, mixer: 5.5 dB, bias tee:  0.2 dB), the reduction in signal due to a non-perfect impedance match (1.5 dB), and the input noise of our lock-in amplifier (3 dB). These noise sources contribute to the degradation of the previously estimated spin sensitivity of the 1 $\Omega$ microcoil. By eliminating these additional noise sources and signal losses from our measurement, we could potentially achieve a sensitivity of $N_{min} \approx 9 \cdot 10^{9}$ spins/Hz\textsuperscript{1/2}, with a remaining discrepancy of 3 from the theoretical evaluation. For example, replacing the directional coupler with a circulator could potentially lead to an 8-fold improvement in SNR.

The remaining discrepancy between theory and experiment could be explained by the degradation of the BPDA sample, the volume estimation of the BPDA sample used, or magnetic field inhomogeneities of the modulation coil or the microresonator. Degradation of the BPDA sample could result in a lower spin density than assumed for our calculation, reducing the effective spin density $N$. The volume estimation $V_s$ is done based on SEM images similar to Fig. \ref{fig:TEM_measurement}a, and is expected to have only a small effect on the estimated spin sensitivity. Inhomogeneities of the magnetic fields and the specific location and size of the sample within an inhomogeneous field, however, can have a significant effect on the ESR signal intensity \cite{fajer1982FieldInhomogeneities, yordanov2002CavityConstruction}.

\end{document}